\documentclass[runningheads]{llncs}
\usepackage{booktabs}
\usepackage[T1]{fontenc}
\usepackage{graphicx}
\usepackage{hyperref}
\usepackage{todonotes}
\begin{document}
\title{Lash 1.0 (System Description)}
\author{Chad E. Brown\inst{1} \and Cezary Kaliszyk\inst{2}\orcidID{0000-0002-8273-6059}}
\authorrunning{C. Brown, C. Kaliszyk}
\institute{Czech Technical University in Prague \and
University of Innsbruck\\
\email{cezary.kaliszyk@uibk.ac.at}}
\maketitle
\begin{abstract}
  Lash is a higher-order automated theorem prover created as a
  fork of the theorem prover Satallax.  The basic underlying calculus
  of Satallax is a ground tableau calculus whose rules
  only use shallow
  information about the terms and formulas taking part in the rule.
  Lash uses new, efficient C representations of vital
  structures and operations. Most importantly, Lash uses a C
  representation of (normal) terms with perfect sharing along with a C
  implementation of normalizing substitutions.
  We describe the ways
  in which Lash differs from Satallax and the performance improvement
  of Lash over Satallax when used with analogous flag settings.
  With a 10s timeout Lash outperforms Satallax on a collection TH0 problems
  from the TPTP.
  We conclude with ideas for continuing the development of Lash.
\end{abstract}
\section{Introduction}

Satallax~\cite{Brown12,Faerber2016} is an automated theorem prover for higher-order logic
that was a top competitor in the THF division of CASC~\cite{CASC-J10} for most of the 2010s.
The basic calculus of Satallax is a complete ground tableau
calculus~\cite{BrownSmolkaEFO,BrownSmolka2010,BackesBrown2011}.
In recent years the top systems of the THF division of CASC are
primarily based on resolution and superposition~\cite{Zipperposition,HOVampire,LeoIII}.
At the moment it is an open question whether there is a research
and development path via which a tableau based prover could
again become competitive.
As a first step towards answering this question
we have created a fork of Satallax, called Lash, focused
on giving efficient C implementations of
data structures and operations needed for search in the basic calculus.

Satallax was partly competitive due to (optional) additions
that went beyond the basic calculus. Three of the most successful
additions were the use of higher-order pattern clauses during search,
the use of higher-order unification as a heuristic to suggest
instantiations at function types
and the use of the first-order theorem prover E as a backend
to try to prove the first-order part of the current state
is already unsatisfiable.
Satallax includes flags that can be used to activate or deactivate such additions
so that search only uses the basic calculus.
They are deactivated by default.
Satallax has three representations of terms in Ocaml.
The basic calculus rules use the primary representation.
Higher-order unification and pattern clauses make use of a representation
that includes a case for metavariables to be instantiated.
Communication with E uses a third representation restricted
to first-order terms and formulas.
When only the basic calculus is used, only the primary representation is needed.

Assuming only the basic calculus is used only limited information
about (normal) terms is needed during the search.
Typically we only need to know the outer structure of the
principal formulas of each rule, and so the full term
does not need to be traversed.
In some cases Satallax either implicitly or explicitly
traverses the term.
The implicit cases are when a rule needs to know if two terms are equal.
In Satallax, Ocaml's equality is used to test
for equality of terms, implicitly relying on a recursion over the term.
The explicit cases are quantifier rules that instantiate with
either a term or a fresh constant. In the former case
we may also need to normalize the result after instantiating
with a term.

In order to give an optimized implementation of the basic calculus
we have created a new theorem prover, Lash\footnote{Lash 1.0 along with accompanying material is available at {\url{http://grid01.ciirc.cvut.cz/~chad/ijcar2022lash/}}.},
by forking a recent version of Satallax (Satallax 3.4), the
last version that won the THF division of CASC (in 2019).
Generally speaking, we have removed all the additional code
that goes beyond the basic calculus.
In particular we do not need terms with metavariables since
we support neither pattern clauses nor higher-order unification in Lash.
Likewise we do not need a special representation for first-order
terms and formulas since Lash does not communicate with E.
We have added efficient C implementations of (normal) terms with perfect sharing.
Additionally we have added new efficient C implementations
of priority queues and the association of formulas
with integers (to communicate with MiniSat).
To measure the speedup given by the new parts of the implementation
we have run Satallax 3.4 using flag settings that only use
the basic calculus and Lash 1.0 using the same flag settings.
We have also compared Lash to Satallax 3.4 using Satallax's default strategy with a timeout of 10s,
and have found that Lash 1.0 outperforms Satallax with this short timeout
even when Satallax is using the optional additions (including calling E).
We describe the changes and present a number of
examples for which the changes lead to a significant speedup.

\section{Preliminaries}

We will presume a familiarity with
simple type theory and only give a quick
description to make our use of notation clear, largely following~\cite{BrownSmolka2010}.
We assume a set of base types,
one of which is the type $o$ of propositions (also called booleans),
and the rest we refer to as sorts.
We use $\alpha,\beta$ to range over sorts and $\sigma,\tau$ to range over types.
The only types other than base types are function types $\sigma\tau$,
which can be thought of as the type of functions from $\sigma$ to $\tau$.

All terms have a unique type and are inductively defined
as (typed) variables, (typed) constants,
well-typed applications $(t~s)$ and $\lambda$-abstractions $(\lambda x.t)$.
We also include the logical constant $\bot$ as a term of type $o$,
terms (of type $o$) of the form $(s\Rightarrow t)$ (implications)
and $(\forall x.t)$ (universal quantifiers) where $s,t$ have type $o$
and terms (of type $o$) of the form $(s =_\sigma t)$ where $s,t$ have a common type $\sigma$.
We also include choice constants $\varepsilon_\sigma$ of $(\sigma o)\sigma$
at each type $\sigma$.
We write $\neg t$ for $t\Rightarrow\bot$ and $(s\not=_\sigma t)$
for $(s =_\sigma t \Rightarrow \bot)$.
We omit type parentheses
and type annotations
except where they are needed for clarity.
Terms of type $o$ are also called propositions. 
We also use $\top$, $\lor, \land, \exists$ with the understanding that
these are notations for equivalent propositions in the set of terms above.

We assume terms are equal if they are the same up to $\alpha$-conversion
of bound variables (using de Bruijn indices in the implementation).
We write $[s]$ for the $\beta\eta$-normal form of $s$.

The tableau calculi of~\cite{BrownSmolka2010} (without choice) and~\cite{BackesBrown2011} (with choice)
define when a branch is refutable. A branch is a finite set of
normal propositions. We let $A$ range over branches and write
$A,s$ for the branch $A\cup\{s\}$.
We will not give a full calculus,
but will instead discuss a few of the rules with surprising properties.
Before doing so we emphasize rules that are {\emph{not}} in the calculus.
There is no cut rule stating that if $A,s$ and $A,\neg s$ are refutable,
then $A$ is refutable. (During search such a rule would require
synthesizing the cut formula $s$.)
There is also no rule stating that if the branch
$A,(s=t),[p s],[p t]$ is refutable,
then $A,(s=t),[p s]$ is refutable (where $s,t$ have type $\sigma$ and $p$ is a term of type $\sigma o$).
That is, there is no rule for rewriting into arbitrarily deep positions using equations.

All the tableau rules only need to examine the outer structure
to test if they apply (when searching backwards for a refutation).
When applying the rule, new formulas are constructed and added to the
branch (or potentially multiple branches, each a subgoal to be refuted).
An example is the confrontation rule, the only rule involving positive
equations.
The confrontation rule states that
if $s=_\alpha t$ and $u\not=_\alpha v$ are on a branch $A$
(where $\alpha$ is a sort), then
we can refute $A$ by refuting $A,s\not=u,t\not=u$
and $A,s\not=v,t\not=v$.
A similar rule is the mating rule, which states that
if $p s_1\ldots s_n$ and $\neg p t_1\ldots t_n$ are on a branch $A$
(where $p$ is a constant of type $\sigma_1\cdots\sigma_n o$),
then we can refute $A$ by refuting each of the branches
$A,s_i\not= t_i$ for each $i\in\{1,\ldots,n\}$.
The mating rule demonstrates how disequations can appear on a branch
even if the original branch to refute contained no reference to equality
at all.
One way a branch can be closed is if $s\not=s$ is on the branch.
In an implementation, this means an equality check
is done for $s$ and $t$ whenever
a disequation $s\not=t$ is added to the branch.
In Satallax this requires Ocaml to traverse the terms.
In Lash this only requires comparing the unique integer ids the implementation
assigns to the terms.

The disequations generated on a branch play an important role.
Terms (of sort $\alpha$) occuring on one side of a disequation on a branch are called
{\emph{discriminating terms}}.
The rule for instantiating a quantified formula $\forall x.t$ (where $x$ has sort $\alpha$)
is restricted to instantiating with discriminating terms
(or a default term if no terms of sort $\alpha$ are discriminating).
During search in Satallax this means there is a finite set of
permitted instantiations (at sort $\alpha$) and this set grows
as disequations are produced. Note that, unlike most
automated theorem provers, the instantiations do not arise from unification.
In Satallax (and Lash) when $\forall x.t$ is being processed
it is instantiated with all previously processed instantiations.
When a new instantiation is produced, previously processed universally
quantified propositions are instantiated with it.
When $\forall x.t$ is instantiated with $s$,
then $[(\lambda x.t) s]$ is added to the branch.
Such an instantiation is the important case where the new
formula involves term traversals: both for substitution and normalization.
In Satallax the substitution and normalization require multiple term
traversals. In Lash we have used normalizing substitutions and memorized
previous computations, minimizing the number of term traversals.
The need to instantiate arises when processing either a universally quantified
proposition (giving a new quantifier to instantiate) or a disequation at a sort (giving new discriminating terms).

We discuss a small example both Satallax and Lash can
easily prove.
We briefly describe what
both do in order to give the flavor of the procedure and (hopefully) prevent
readers from assuming the provers behave too similarly from readers based on
other calculi (e.g., resolution).

Example {\verb+SEV241^5+} from TPTP v7.5.0~\cite{TPTP}
({\verb+X5201A+} from {\sc{Tps}}~\cite{AndrewsBINPX96})
contains a minor amount of features going beyond first-order logic.
The statement to prove is
$$\forall x. U~x \land W~x \Rightarrow \forall S. (S = U \lor S = W) \Rightarrow S x.$$
Here $U$ and $W$ are constants of type $\alpha  o$,
$x$ is a variable of type $\alpha$
and $S$ is a variable of type $\alpha o$.
The higher-order aspects of this problem are the quantifier for $S$ (though this
could be circumvented by making $S$ a constant like $U$ and $W$)
and the equations between predicates (though these could be circumvented by
replacing $S = U$ by $\forall y. S y \Leftrightarrow U y$
and replacing $S = W$ similarly). The tableau rules effectively do
both during search.

Satallax never clausifies. The formula above is negated and assumed.
We will informally describe tableau rules as splitting the problem into subgoals,
though this is technically mediated through MiniSat
(where the set of MiniSat clauses is unsatisfiable when all branches are closed).
Tableau rules are applied until the problem involves a constant $c$ (for $x$), a constant $S'$ for $S$
and assumptions $U~c$, $W~c$, $S' = U \lor S' = W$ and $\neg S' c$ on the branch.
The disjunction is internally $S'\not= U \Rightarrow S' = W$ and the implication rule
splits the problem into two branches, one with $S' = U$ and one with $S' = W$.
Both branches are solved in analogous ways and we only describe the $S'=U$ branch.
Since $S'=U$ is an equation at function type, the relevant rule
adds $\forall y.S' y = U y$ to the branch.
Since there are no disequations on the branch, there is no instantiation
available for $\forall y.S' y = U y$. In such a case, a default instantiation
is created and used. That is, a default constant $d$ (of sort $\alpha$) is generated and
we instantiate with this $d$, giving $S' d =_o U d$.
The rule for equations at type $o$ splits into two subgoals: one branch with $S' d$ and $U d$
and another with $\neg S' d$ and $\neg U d$.
On the first branch we mate $S' d$ with $\neg S' c$ adding the disequation $d\not= c$
to the branch. This makes $c$ available as an instantiation for 
$\forall y.S' y = U y$. After instantiating with $c$ the rest of the subcase is straightforward.
In the other subgoal we mate $U~c$ with $\neg U d$ giving the disequation $c\not= d$.
Again, $c$ becomes available as an instantiation and the rest of the subcase is straightforward.

\section{Terms with Perfect Sharing}

Lash represents normal terms as C structures, with a unique integer id
assigned to each term.
The structure contains a tag indicating which kind of term
is represented, a number that is used to either indicate
the de Bruijn index (for a variable), the name (for a constant),
or the type (for a $\lambda$-abstraction, a universal quantifier,
a choice operator, or an equation).
Two pointers (optionally) point to relevant subterms in each case.
In addition the structure maintains the information of which de Bruijn
indices are free in the term (with de Bruijn indices limited to a maximum of 255).
Knowing the free de Bruijn indices of terms makes recognizing potential $\eta$-redexes
possible without traversing the $\lambda$-abstraction.
Likewise it is possible to determine when shifting and substitution of de Bruijn indices
would not affect a term, avoiding the need to traverse the term.

In Ocaml only the unique integer id is directly revealed and this
is sufficient to test for equality of terms.
Hash tables are used to uniquely assign types to integers
and strings (for names) to integers and these integers
are used to interface with the C code.
Various functions are used in the Ocaml-C interface
to request the construction of (normal) terms.
For example, given the two Ocaml integer ids $i$ and $j$ corresponding to terms
$s$ and $t$, the function {\tt{mk\_norm\_ap}} given $i$ and $j$
will return an integer $k$
corresponding to the normal term $[s~t]$. The C implementation recognizes
if $s$ is a $\lambda$-abstraction and performs all $\beta\eta$-reductions
to obtain a normal term. Additionally, the C implementation treats terms
as graphs with perfect sharing, and additionally caches previous
operations (including substitutions and de Bruijn shifting) to prevent
recomputation.

In addition to the low-level C term reimplementation, we have also provided
a number of other low-level functionalities replacing the slower parts of
the Ocaml code. This includes low-level priority queues, as well as C code
used to associate the integers representing normal propositions with integers
that are used to communicate with MiniSat.
The MiniSat integers are nonzero and satisfy the property
that minus on integers corresponds to negation of propositions.

\section{Results and Examples}

The first mode in the default schedule for Satallax 3.1 is {\sc{mode213}}.
This mode activates one feature that goes beyond the basic calculus:
pattern clauses. Additionally the mode sets a flag that tries
to split the initial goal into several independent subgoals before
beginning the search proper.
Through experimentation we have found that setting a flag (common to both Satallax and Lash)
to essentially prevent MiniSat from searching (i.e., only using MiniSat to
recognize contradictions that are evident without search)
often improves the performance.
We have created a modified mode {\sc{mode213d}}
that deactivates these additions (and delays the use of MiniSat)
so that Satallax and Lash will
have a similar (and often the same) search space.
(Sometimes the search spaces differ due to differences in the
way Satallax and Lash enumerate instantiations for function types,
an issue we will not focus on here.)
We have also run Lash with many variants of Satallax modes with similar modifications.
From such test runs we have created a 10 second schedule consisting of 5 modes.

To give a general comparison of Satallax and Lash
we have run both 
on 2053 TH0 problems from a recent release
of the TPTP~\cite{TPTP} (7.5.0).
We initially selected all problems
with TPTP status of Theorem or Unsatisfiable (so they should be
provable in principle) without polymorphism (or similar
extensions of TH0).
We additionally removed a few problems that could not be parsed by Satallax 3.4
and removed a few hundred problems big enough to activate SINE in Satallax 3.4.

We ran Lash for 10 seconds with its default schedule over this problem set.
For comparison, we have run Satallax 3.4 for 10s in three different ways:
using the Lash schedule (since the flag settings make sense for both systems)
and
using Satallax 3.4's default schedule both with and without access to E~\cite{EHO}.
The results are reported in Table~\ref{tab:res}.
It is already promising that Lash has the ability to slightly outperform
Satallax even when Satallax is allowed to call E.

\begin{table}[tb]
  \caption{\label{tab:res}Lash vs. Satallax on 2053 TH0 Problems.}
  \centering
\begin{tabular}{lc}
\toprule
Prover & Problems Solved \\\midrule

Lash & 1501 (73\%) \\
Satallax (with E) & 1487 (72\%) \\
Satallax (without E) & 1445 (70\%) \\
Satallax (Lash Schedule) & 1412 (69\%) \\
\bottomrule
\end{tabular}
\vspace{-3mm}
\end{table}

To get a clearer view of the improvement we discuss a few
specific examples.

TPTP problem {\verb+NUM638^1+}
(part of Theorem 3 from the {\sc{Automath}} formalization of Landau's book)
is about the natural numbers (starting from 1).
The problem assumes a successor function $s$ is injective and
that every number other than $1$ has a predecessor.
An abstract notion of existence is used by having a constant
${\mathsf{some}}$ of type $(\iota o)o$ about which no extra assumptions are made,
so the assumption is formally
$\forall x. x\not= 1 \Rightarrow {\mathsf{some}} (\lambda u. x = s u)$.
For a fixed $n$, $n\not=1$ is assumed and the conjecture
to prove is the negation of the implication
$(\forall x y.n=s x \Rightarrow n=s y \Rightarrow x = y)\Rightarrow \neg(\mathsf{some} (\lambda u. n = s u))$.
The implication is assumed and the search must rule out the
negation of the antecedent (i.e., that $n$ has two predecessors)
and the succedent (that $n$ has no predecessor).
Satallax and Lash both take 3911 steps to prove this example.
With {\sc{mode213d}},
Lash completes the search in 0.4s while Satallax requires almost 29s.

TPTP problem {\verb+SEV108^5+} ({\verb+SIX_THEOREM+} from {\sc{Tps}}~\cite{AndrewsBINPX96})
corresponds to proving the Ramsey number R(3,3) is at most 6.
The problem assumes there is a symmetric binary relation
$R$ (the edge relation of a graph with the sort as vertices)
and there are (at least) 6 distinct elements.
The conclusion is that there are either 3 distinct elements
all of which are $R$-related or 3 distinct elements
none of which are $R$-related.
Satallax and Lash can solve the problem in 14129 steps with mode {\sc{mode213d}}.
Satallax proves the theorem in 0.153 seconds while
Lash proves the theorem in the same number of steps but in 0.046 seconds.

The difference is more impressive if we consider the modified
problem of proving R(3,4) is at most 9. That is, we assume
there are (at least) 9 distinct elements and modify
the second disjunct of the conclusion to be that there
are 4 distinct elements none of which are $R$-related.
Satallax and Lash both use 186127 steps to find the proof.
For Satallax this takes 44s while for Lash this takes 5.5s.

The TPTP problem {\verb+SYO506^1+} is about an if-then-else operator.
The problem has a constant $c$ of type $o\iota\iota\iota$.
Instead of giving axioms indicating $c$ behaves as an if-then-else operator,
the conjecture is given as a disjunction:
$$(\forall x y.c~(x=y)~x~y~=~y) \lor \neg(\forall x y.c~\top~x~y~=~x)\lor \neg(\forall x y.c~\bot~x~y~=~y).$$
After negating the conjecture and applying the first few tableau rules the branch will
contain the propositions $\forall x y.c~\top~x~y~=~x$,
$\forall x y.c~\bot~x~y~=~y$
and the disequation $c~(d=e)~d~e\not= e$ for fresh $d$ and $e$ of type $\iota$.
In principle the rules for if-then-else given in~\cite{BackesBrown2011}
could be used to solve the problem without using the universally quantified
formulas (other than to justify that $c$ {\emph{is}} an if-then-else operator).
However, these are not implemented in Satallax or Lash.
Instead search proceeds as usual via the basic underlying procedure.
Both Satallax and Lash can prove the example using modes {\sc{mode0c1}}
in 32704 steps. Satallax performs the search in 9.8 seconds
while Lash completes the search in 0.2 seconds.

In addition to the examples considered above, we have constructed
a family of examples intended to demonstrate the power of the shared
term representation and caching of operations.
Let ${\mathsf{cons}}$ have type $\iota\iota\iota$ and ${\mathsf{nil}}$ have
type $\iota$. For each natural number $n$, consider the proposition $C^n$ given by
$$\overline{n}~(\lambda x.{\mathsf{cons}}~x~x)~({\mathsf{cons}}~{\mathsf{nil}}~{\mathsf{nil}})
= {\mathsf{cons}}~(\overline{n}~(\lambda x.{\mathsf{cons}}~x~x)~{\mathsf{nil}})~(\overline{n}~(\lambda x.{\mathsf{cons}}~x~x)~{\mathsf{nil}})$$
where $\overline{n}$ is the appropriately typed Church numeral.
Proving the proposition $C^n$ does not require any search and merely requires
the prover to normalize the conjecture and note the two sides have the same normal form.
However, this normal form on both sides will be a complete binary tree of depth $n+1$.
We have run Lash and Satallax on $C^{n}$
with $n\in\{20,21,22,23,24\}$ using mode {\sc{mode213d}}.
Lash solves all five problems in the same amount of time, less than 0.02 seconds for each.
Satallax takes $4$ seconds, $8$ seconds, $16$ seconds, $32$ seconds and $64$ seconds.
As expected, since Satallax is not using a shared representation, the computation time
exponentially increases with respect to $n$.

\section{Conclusion and Future Work}

We have used Lash as a vehicle to demonstrate that
giving a more efficient implementation of the underlying
tableau calculus of Satallax can lead to significant
performance improvements.
An obvious possible extension of Lash would be to implement
pattern clauses, higher-order unification and the ability to call E.
While we may do this, our current plans are to focus on
directions that further diverge from the development path
followed by Satallax.

Interesting theoretical work would be to modify
the underlying calculus (while maintaining completeness).
For example the rules of the calculus might be able to
be further restricted based on orderings of ground terms.
On the other hand, new rules might be added
to support a variety of constants with special properties.
This was already done for constants that
satisfy axioms indicating the constant is a
choice, description or if-then-else operator~\cite{BackesBrown2011}.
Suppose a constant $r$ of type $\iota\iota o$ is known to be reflexive
due to a formula $\forall x.r~x~x$ being on the branch.
One could avoid ever instantiating this universally quantified formula
by simply including a tableau rule that extends a branch with $s\not= t$
whenever $\lnot r~s~t$ is on the branch. Similar rules could operationalize
other special cases of universally quantified formulas, e.g., formulas
giving symmetry or transitivity of a relation.
A modification of the usual completeness proof would be required to
prove completeness of the calculus with these additional rules
(and with the restriction disallowing instantiating the corresponding universally
quantified formulas).

Finally the C representation of terms could be extended to include
precomputed special features. Just as the current implementation
knows which de Bruijns are free in the term (without traversing the term),
a future implementation could know other features of the term
without requiring traversal. Such features could be used to guide the search.

\subsubsection{Acknowledgements}

The results were supported by the Ministry of Education, Youth and Sports within the dedicated program ERC CZ under the project POSTMAN no.~LL1902 and the ERC starting grant no.~714034 SMART.


\end{document}